\documentclass[aps,prl,twocolumn,groupedaddress]{revtex4-1}

\usepackage{graphics}
\usepackage{epsfig}
\usepackage{color}
\usepackage{amsmath}
\usepackage{bm}

\begin{document}



\title{Emergence of a higher energy structure in strong field ionization with inhomogeneous electric fields}

\author{L. Ortmann$^{1}$}
\thanks {The first two authors contributed equally}
\email[]{\newline ortmann@pks.mpg.de}
\author{J. A. P\'erez-Hern\'andez$^{2}$}
\email[]{japerez@clpu.es}
\author{M. F. Ciappina$^{3,4}$}
\author{A. Chac\'on$^{5}$}
\author{G. Zeraouli$^{2}$}
\author{M. F. Kling$^{3,6}$}
\author{L. Roso$^{2}$}
\author{M. Lewenstein$^{5,7}$}
\author{A. S. Landsman$^{1}$}
\email[]{landsman@pks.mpg.de}
\affiliation{$^1$Max Planck Institute for the Physics of Complex Systems, N\"othnitzer Stra{\ss}e 38, D-01187 Dresden, Germany}
\affiliation{$^2$Centro de L\'aseres Pulsados (CLPU), Parque Cient\'ifico, E-37008 Villamayor, Salamanca, Spain}
\affiliation{$^3$Max-Planck Institut f\"ur Quantenoptik, Hans-Kopfermann-Str.~1, D-85748 Garching, Germany}
\affiliation{$^4$Institute of Physics of the ASCR, ELI-Beamlines, Na Slovance 2, 182 21 Prague, Czech Republic}
\affiliation{$^5$ICFO-Institut de Ci\`ences Fot\`oniques, The Barcelona Institute of Science and Technology, 08860 Castelldefels (Barcelona), Spain}
\affiliation{$^6$Department f\"ur Physik, Ludwig-Maximilians-Universit\"at M\"unchen, Am Coulombwall 1, D-85748 Garching, Germany}
\affiliation{$^7$ICREA-Instituci\'o Catalana de Recerca i Estudis Avan\c{c}ats, Lluis Companys 23, 08010 Barcelona, Spain}

\date{\today}

\begin{abstract}
Studies of strong field ionization have historically relied on the strong field approximation, which neglects all spatial dependence in the forces experienced by the electron after ionization.  More recently, the small spatial inhomogeneity introduced by the long-range Coulomb potential has been linked to a number of important features in the photoelectron spectrum, such as Coulomb asymmetry, Coulomb focusing, and the low energy structure (LES).  Here, we demonstrate by combined quantum and classical simulations that a small time-varying spatial dependence in the laser electric field creates a prominent higher energy peak at energies above the ``classical cut-off" for direct electrons.  This higher energy structure (HES) originates from direct electrons ionized near the peak of a single half-cycle of the laser pulse. The HES is separated from all other ionization events (providing sub-cycle resolution) and is highly sensitive to the carrier envelope phase (CEP).  The large accumulation of electrons with tuneable energy suggests a promising method for creating a localized source of electron pulses of attosecond duration using  {\em {tabletop laser}} technology.  

\end{abstract}
 
%
\maketitle

When the photon energy of light is many times smaller than the ionization potential of an atom, the ionization occurs either via a tunnel or via a multi-photon ionization process \cite{pazourek,landsmanPhysRep,agostini1979,grasbon_03}.  These two regimes are well distinguished by the Keldysh parameter $\gamma = \sqrt{I_p/2 U_p}$, where $I_p$ is the ionization potential and $U_p$ is the ponderomotive energy of an electron in a laser field \cite{keldysh}.  For $\gamma \leq 1$, the ionization process is dominated by tunneling, whereby the electric field of the laser bends the binding potential of the atom, forming a barrier through which the electron tunnels out and is subsequently accelerated by the strong laser field \cite{corkum}.  
Tunnel ionization  underlies the creation of attosecond pulses via the process of high harmonic generation (HHG) \cite{corkum,anne,krausz_09,lewenstein94A}, as well as a variety of other important applications, including photoelectron holography \cite{huismans,bandrauk}, tomographic imaging of molecular orbitals \cite{Itatani} and electron diffraction \cite{Meckel,blagaDiff,JensDiff}.

Wavelengths used in tunnel ionization experiments are typically in the infrared range (usually around $800$ nm), and have more recently been extended into the mid-IR regime \cite{blaga,JensMidIR1,dura_13,JensMidIR2}. Under these conditions, the laser field is well-described by the dipole approximation, resulting in spatially homogeneous time-varying electric fields.  
The strong field approximation (SFA) \cite{keldysh,faisal,reiss}, which in its standard form neglects the remaining Coulomb force on the ionized electron, has been the dominant tool for investigating electron dynamics under these circumstances.  However, the small spatial dependence introduced by the $1/r$ Coulomb potential has led to a number of interesting phenomena, such as Coulomb asymmetry \cite{bandrauk2,landsman2013} and Coulomb focusing \cite{landsman2013,brabec}.  Of particular interest is the discovery of the low energy structure (LES) using mid-IR pulses \cite{blaga}.  This surprising finding stimulated a great amount of experimental \cite{quan,dura_13,wu,wolter_14,moller_14} and theoretical work \cite{wolter_14,moller_14,liu,yan,xia,rost}, and highlighted the dramatic impact that even a small spatial inhomogeneity in force can have on electron dynamics after strong field ionization.  

At the same time, there has been significant interest in strong field ionization phenomena in the vicinity of nanostructures \cite{peternature,kling,peterprl2006,peterprl2010,peterjpbreview,ropers,z11,sussmann2015field,forg2016attosecond,kimNature,sivis}.  A key characteristic of nanostructures is evanescent near-fields, which can show field-enhancement resulting in a time-dependent spatial inhomogeneity in the presence of a laser pulse. Prior theoretical work also investigated HHG from atoms in inhomogeneous electric fields \cite{kimNature,sivis,husakou,ciappi2012,choi,yavuz,ciappi_opt,ciappiati1d,tahirsfa,tahirJMO, LPL, PRA_}.
It is well-known that HHG yield is much lower than electron yield, since only a tiny fraction of ionized electrons recombine to emit high harmonics.  Hence, while the nanoscopic volume may prohibit an efficient high harmonic conversion \cite{sivis}, efficient generation of electrons from gas targets in the vicinity of a nanostructure appears quite feasible.



In this Letter, we investigate the impact on an electron wavepacket of a time-dependent spatial inhomogeneity, such as the one created when a laser pulse interacts with a nanostructure.  Our approach combines the solution of a three-dimensional time-dependent Schr\"odinger equation (3D-TDSE) with classical trajectory Monte Carlo (CTMC) simulations.  
We find that even a small inhomogeneity can have a dramatic impact, resulting in a prominent higher energy peak, with energy above the classical cut-off of $2 U_p$ for direct electrons in homogeneous laser fields. 
As CTMC simulations indicate, the electrons comprising this HES are all direct electrons, rather than scattered electrons normally associated with higher energies \cite{corkum, blaga, lewenstein1995rings,salieres2001feynman, suarez2015above}.  
A small time-dependent spatial inhomogeneity is introduced along the direction of polarization of the laser.  
Such inhomogeneity can occur in the vicinity of a nanostructure, where it is created by the localized field enhancement \cite{husakou, choi, ciappi2012}.  The electric field is then given by:
\begin{equation}
{\bf E}(z,t)=E_0 (1+2 \beta z) f(t) \cos(\omega t+ \varphi) \hat{z}
\label{Efield}
\end{equation}
where $E_0$ is the electric field amplitude in atomic units ($E_0=\sqrt{I/I_0}$ with $I_0=3.5\times 10^{16}$ W/cm$^2$), $\omega=0.057$ a.u.  (corresponding to a wavelength of $\lambda=800$ nm), $\hat{z}$ is the direction of polarization, and $\varphi$ is the carrier-envelope phase (CEP).  The sine-squared envelope is given by: $f(t)=\cos^{2}\left(\frac{\omega t}{2N }\right)$, where $N$ is a number of cycles in a pulse.  The parameter $\beta$ defines
the `strength'  of the inhomogeneity and has units of inverse length \cite{husakou,yavuz}. 
 
To calculate the energy-resolved photoelectron spectra, $P(E)$, we solve the 3D-TDSE in the length gauge for hydrogen and helium atoms:
\begin{equation}
\label{tdse}
\frac{i\partial \Psi({\bf{r}},t)}{\partial t}=H\Psi({\bf{r}},t)=\left [-\frac{\nabla^{2}}{2} + V(r) +V_l({{\bf{r}},t})\right ]\Psi({\bf{r}},t)\\
\end{equation}
where $V(r)$ is the atomic potential, given by $-1/r$ for hydrogen (the analytic expression for helium is given in ~\cite{tong}), and $V_{l}(\mathbf{r},t)= - \int ^\mathbf{r} d\mathbf{r'}\cdot\mathbf{E}(\mathbf{r'},t)$ represents interaction with the laser field.
The time-dependent electronic wave function, $\Psi({\bf{r}},t)$, can be expanded in spherical harmonics:
\begin{eqnarray}
\label{spherical}
\Psi({\bf{r}},t)&=&\Psi({r,\theta, \phi},t)\approx\sum_{l=0}^{L-1}\sum_{m=-l}^{l}\frac{\Phi_{lm}(r,t)}{r}Y_{l}^{m}(\theta,\phi)
\end{eqnarray}
where $L$ is the number of partial waves, which depends on laser parameters. Here, we use values up to $L\approx300$,  to avoid spurious effects in the photoelectron spectrum due to relatively high laser intensity.  Since the laser field is linearly polarized, only $m=0$ magnetic quantum numbers are considered. 


 We assume that before switching on the laser, the target atoms hydrogen and helium are
in their $1s$ and $2s$ ground states, respectively.  
Inserting Eq.~(\ref{spherical}) into Eq.~(\ref{tdse}) and using $\cos \theta Y_{l}^{0}=c_{l-1}Y_{l-1}^{0}+c_{l}Y_{l+1}^{0}$ and $\cos^{2} \theta Y_{l}^{0}=c_{l-2}c_{l-1}Y_{l-1}^{0}+(c_{l-1}^{2}+c_{l}^{2})Y_{l}^{0}+c_{l}c_{l+1}Y_{l+2}^{0}$, where $c_{l}=\sqrt{(l+1)^2/(2l+1)(2l+3)}$, we arrive at the following coupled differential equations for the radial part part of the wavefunction:
\begin{eqnarray}
\label{diag}
i\frac{\partial\Phi_{l}}{\partial{t}}=\left [-\frac{1}{2}\frac{\partial^{2}}{\partial r^{2}}+\frac{l(l+1)}{2r^2}-\frac{1}{2} \right ]\Phi_{l}\nonumber\\
+\beta r^{2}E(t)\left(c_{l}^{2}+c_{l-1}^{2}\right)\Phi_{l}\nonumber\\
+r E(t)\left(c_{l-1}\Phi_{l-1}+c_{l}\Phi_{l+1}\right)\nonumber\\
+\beta r^{2}E(t)\left(c_{l-2}c_{l-1}\Phi_{l-2}+c_{l}c_{l+1}\Phi_{l+2}\right).
\end{eqnarray}
Equation (\ref{diag}) is solved using the Crank-Nicolson algorithm.  To calculate the energy-resolved photoelectron spectra $P(E)$ and the two-dimensional electron distributions $\mathcal{H}(p,\theta)$, we use the window function approach developed by Schafer and colleagues ~\cite{schaferwop1,schaferwop,schaferwop2}.
\begin{figure}
\resizebox{3in}{!}{\includegraphics[angle=0]{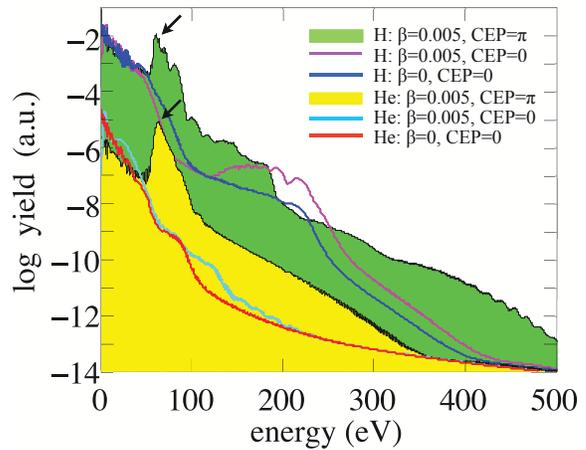}}  
\caption{3D TDSE photoelectron spectra in hydrogen and helium atoms generated by laser pulses described in Eq. (\ref{Efield}) for different values of ${\beta}$ parameter and for two values of the CEP.  The laser intensity is $I=4.496\times 10^{14}$ W/cm$^2$ (or $E_0 = 0.1132$ a.u.), and $N=5$.
Two arrows show a prominent HES appearing for non-homogeneous fields at $CEP=\pi$.}
\label{fig1}
\end{figure}

Figure \ref{fig1} shows the ATI spectra for hydrogen and helium atoms calculated using 3D-TDSE.  As expected, the electron yield from helium is significantly lower than from hydrogen due to a higher ionization potential.  For both atoms, when $CEP = \pi$, the yield is significantly enhanced by the presence of the spatial inhomogeneity.  

Importantly, a prominent HES appears above $60$ eV for $CEP=\pi$, but not for $CEP=0$, indicating a break in the field inversion symmetry.   
Note that the electrons comprising the HES are relatively high in energy, beyond the classical cutoff of $2 U_p$ observed for direct electrons \cite{blaga}.  To understand the physical origin of HES, we use CTMC simulations to investigate electron trajectories after ionization of helium.
Here,  single trajectories are launched at a starting phase $\varphi_0 = \omega \cdot t_0$, with velocity $v_\perp$ perpendicular to the laser polarization direction. The probability distribution at the tunnel exit is given by the Ammosov-Delone-Krainov (ADK) formula ~\cite{ammosov1986tunnel,delone1991energy}, typically used to model strong field ionization \cite{arissian,rydberg,JPhysBTime,eichmann}:
\begin{align}
\begin{split}
 P(t_0,v_\perp) = \exp &\left(-\frac{2(2I_p(t_0))^{3/2}}{3 E(t_0)}\right)  \\ &\cdot \exp\left(-\frac{v_\perp^2 \sqrt{2 I_p(t_0)}}{E(t_0)}\right), \label{eq:ADK}
\end{split}
 \end{align}
where the laser field $E(t_0)$ is given by Eq. (\ref{Efield}) with $z=0$, corresponding to an atom centered at the origin. $I_p$ denotes the Stark shifted ionization potential \cite{{hofmann2013comparison}}
\begin{equation}
  I_p(E(t_0))=I_{p,0}+\frac{1}{2}(\alpha_N-\alpha_I)F(t_0)^2 , \label{eq:starkShiftedIp}
\end{equation}
with $\alpha_N$ and $\alpha_I$ representing the polarizability of the atom and ion, respectively.  The tunnel exit radius is obtained using parabolic coordinates \cite{pfeiffer2012attoclock,hofmann2013comparison,Cornelia2,Cornelia3}.
The dynamics of each electronic trajectory after ionization is solved numerically by integrating the Newton's equations following the method in ~\cite{pfeiffer2012attoclock}, which takes into account the laser field, the Coulomb potential and the induced dipole (the latter is negligible in helium).
\begin{figure}[here]
\resizebox{3.4in}{!}{\includegraphics[angle=270]{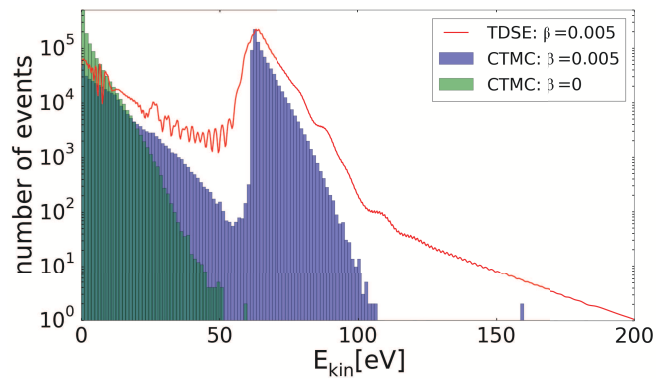}} 
\caption{Histogram of the electron yield as a function of the final kinetic energy obtained in a CTMC calculation for helium at CEP=$\pi$ with the parameters given in Fig. \ref{fig1} for two different values of $\beta$, where $\beta=0$ corresponds to the homogeneous field.}
\label{fig2}
\end{figure}
Figure \ref{fig2} shows electron yield as a function of energy obtained with CTMC simulations.  As can be seen, the prominent higher energy peak (starting around 60 eV) observed in 3D-TDSE (red curve) is well-reproduced.  The oscillations in the TDSE curve observed at lower electron energies are due to inter-cycle interference, which is not captured by classical trajectory simulations. 
Also, the high energetic tail close to the cut-off is missing in the CTMC results, since this feature is due to rescattering events which were not included in our classical calculations \cite{yang1993intensity,paulus_94}.
The higher energy peak is more pronounced in CTMC compared to TDSE simulations.  This is likely because, as we show below, this peak originates from electrons ionized near the maximum of the laser field.  The ADK distribution used in CTMC simulations is known to over-estimate the relative probability of ionization near the laser field maxima ~\cite{yudin2001nonadiabatic}, hence making the peak more pronounced than what is observed in TDSE simulations (see Fig. \ref{fig2}).


\begin{figure}[here]
\resizebox{3.2in}{!}{\includegraphics[angle=0]{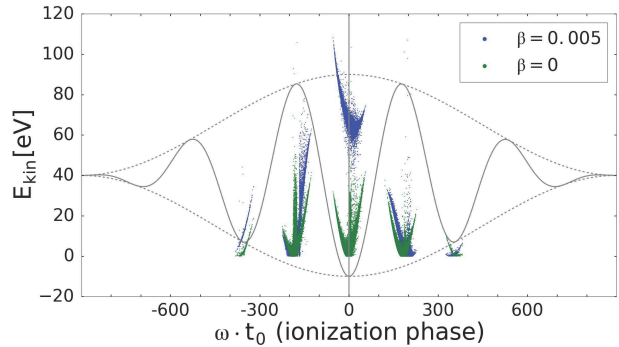}} 
\caption{Kinetic energy of electrons as a function of ionization phase from a CTMC calculation for helium at CEP=$\pi$ for homogeneous ($\beta=0$) and non-homogeneous ($\beta=0.005$) electric fields.  The corresponding laser field is plotted in grey.  All laser parameters are the same as in Fig. \ref{fig2}. }
\label{fig2b}
\end{figure}
Figure \ref{fig2b} establishes the physical origin of the HES by comparing the final electron kinetic energy as a function of ionization phase for homogeneous and inhomogeneous fields.  By far, the most dramatic influence of the spatial inhomogeneity occurs in the central cycle, corresponding to the maximum probability of ionization along the direction of increasing field. 
As Fig. \ref{fig2b} shows, field inhomogeneity causes electrons ionized near the laser field maximum to get accelerated to over 60 eV, whereas these same electrons have much smaller energies in homogeneous fields.  In fact, the electrons ionized near the peak by homogeneous fields are known to have low final energies (see also Fig. \ref{fig2b}), thereby contributing to Rydberg states \cite{rydberg} and the zero energy structure \cite{JensMidIR2,wolter_14}. 
A closer look at the figure reveals that the large accumulation of trajectories in the range of 60 to 65 eV stems primarily from electrons ionized just after the peak of the laser field, whereas the higher energy electrons (above 70 eV) come from ionization before the laser field maximum.  

Since all electrons in the HES come from a single half-cycle, they are distinctly separated in energy from all other ionization events, suggesting that inter-cycle interference should only be observed at lower electron energies. This is in fact supported by TDSE simulations (see Fig. \ref{fig2}, red curve), which show significant oscillations (indicative of inter-cycle interference) only before the higher energy peak.  

\begin{figure}[here]
\resizebox{3.5in}{!}{\includegraphics[angle=0]{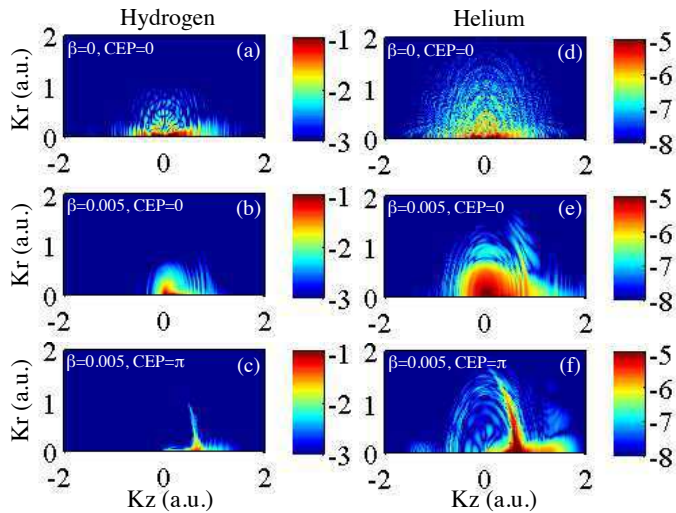}}
\caption{Two-dimensional electron momentum distributions ($k_z,k_r$) using the exact 3D-TDSE calculation for hydrogen (a), (b), (c) and for helium (d), (e) and (f) respectively, for the same laser parameters used in Fig.~\ref{fig1}.}
\label{fig4}
\end{figure}

Based on the above analysis, the appearance of a HES should coincide with a depletion of low energy electrons, which get accelerated by the field inhomogeneity.  This depletion can be clearly observed in 3D-TDSE simulations showing electron momenta distributions for hydrogen, Fig. \ref{fig4}(a)-(c), and helium, Fig. \ref{fig4}(d)-(f), for homogeneous and non-homogeneous electric fields.  The pronounced asymmetric structure in Fig. \ref{fig4}(f) corresponds to the higher energy asymmetric curve shown in blue in the middle of Fig. \ref{fig2b}.
 In agreement with our CTMC simulations, the high energy electrons comprising the asymmetric structure in Fig. \ref{fig4}(f) come from before the peak of the laser field, resulting in a positive final momentum. 
To further investigate the HES as a function of experimental parameters, we performed CTMC simulations for four different CEP phases, corresponding to CEP $=(0, \pi/2, \pi, 3 \pi/2)$, at a fixed laser intensity of $4.5 \cdot 10^{14} W/cm^2$.  We also varied the field intensity in the range of $3-6 \cdot 10^{14} W/cm^2$, corresponding to $E_0 =0.025-0.1308$ a.u.  We find that the higher energy structure occurs at all values of CEP, except for $CEP=0$.  
In all cases, the electrons forming the higher energy peak come from within a single laser half-cycle, which ionizes in the direction of increasing field.
Moreover, for CEP=$\pi/2$, the higher energy peak becomes sharper and more confined, while the opposite is true for CEP=$3 \pi/2$.  This suggests the HES can be a good complement to the current method of CEP characterization, which relies on rescattered electrons \cite{paulusCEP,milosevic_03}.

Increasing the laser intensity broadens the peak and shifts it to higher energies.  
For all intensities, the electrons comprising the peak had energies above the classical cut-off of $2 U_p$ for direct electrons in spatially homogeneous laser fields.



In conclusion, using laser fields with a weak spatial inhomogeneity, we find a substantial enhancement of a few orders of magnitude in the efficiency of the photoelectron spectrum.  This enhancement corresponds to a formation of a prominent higher energy structure above the classical cut-off for direct electrons, and a concomitant suppression of low energy electrons.  The electrons comprising the HES come from ionization within a single half-cycle of the electric field.  Note that this is in sharp contrast to the typical situation in strong field ionization of atomic gas, where neighboring cycles of comparable amplitude make similar contributions to the total electron spectrum  \cite{carpetPRL}.    Hence, the field inhomogeneity leads to a sub-cycle resolved HES, without the need to use half-cycle light transients \cite{wirth}.  
Finally, the fact that the prominent higher energy peak comes from a narrow time window, well within a single half-cycle of the laser pulse, may be used to create localized sources of monoenergetic electron beams of sub-femtosecond duration.
Such sources would take the techniques of classical electron diffraction into the attosecond domain, enabling the investigation of dynamic changes of electron distribution in complex systems, such as nanostructures and biological molecules \cite{fill,baum}.

\begin{acknowledgments}
 J. A. P.-H. and L. R. acknowledge support from Spanish Ministerio de Econom\'{\i}a y Competitividad through the 
FURIAM Project No. FIS2013-47741-R and LaserLab IV Grant  Agreement No. 654148.  A.S.L. is supported by the Max Planck Center for Attosecond Science (MPC-AS). 
M. C. was supported by the project ELI-Extreme Light Infrastructure-phase 2
(CZ.02.1.01/0.0/0.0/15 008/0000162 ) from European Regional Development Fund.
A. C. and M. L. acknowledge support from ERC AdG OSYRIS, Spanish MINECO (FIS2013-46768-P FOQUS and Severo Ochoa SEV-2015-0522), Catalan Agaur SGR 874 and Fundaci\'o Cellex. 
M. F. K. is grateful for support by the EU via the ERC grant ATTOCO (no. 307203) and by the DFG via the excellence center "Munich Centre for Advanced Photonics".
\end{acknowledgments}



\begin{thebibliography}{}

\bibitem{pazourek} R. Pazourek, S. Nagele, and J. Burgd\"orfer, Rev. Mod. Phys. {\bf 87}, 765 (2015).
\bibitem{landsmanPhysRep} A.S. Landsman and U. Keller, Phys. Rep. {\bf 547}, (2015).
\bibitem{agostini1979} P. Agostini, F. Fabre, G. Mainfray, G. Petite, and N. K. Rahman, Phys. Rev. Lett. {\bf 42}, 1127 (1979).
\bibitem{grasbon_03} F. Grasbon, G. G. Paulus, H. Walther, P. Villoresi, G. Sansone, S. Stagira, M. Nisoli, and S. De Silvestri, Phys. Rev. Lett. {\bf 91}, 173003 (2003).
\bibitem{keldysh} L.V. Keldysh, J. Exp. Theor. Phys. {\bf 20}, 1307 (1965).
\bibitem{corkum} P.B. Corkum, Phys. Rev. Lett. {\bf 71}, (1993).
\bibitem{anne} M. Ferray, A. L'Hullier, X.F. Li, L.A. Lompre, G. Mainfray, and C. Manus, J. Phys. B {\bf 21}, L31 (1988).
\bibitem{krausz_09}  F. Krausz and M. Ivanov, Rev. Mod. Phys. {\bf 81}, 163 (2009).
\bibitem{lewenstein94A} M. Lewenstein, Ph. Balcou, M. Y. Ivanov,  A. L'Hullier, and P. B. Corkum, Phys. Rev. A {\bf 49}, 2117 (1994).
\bibitem{huismans} Y. Huismans et al., Science {\bf 331}, 61 (2011).
\bibitem{bandrauk} X.B. Bian and A.D. Bandrauk, Phys. Rev. Lett. {\bf 108}, 263003 (2012).
\bibitem{Itatani} J. Itatani, J. Levesque, D. Zeidler, H. Niikura, H. Pepin, J.C. Keiffer, P.B. Corkum, and D.M. Villeneuve, Nature {\bf 432}, 867 (2004).
\bibitem{Meckel} M. Meckel, et al., Science {\bf 320}, 1478 (2008).
\bibitem{blagaDiff} C.I. Blaga, J.L. Xu, A.D. DiChiara, E. Sistrunk, K. Zhang, P. Agostini, T.A. Miller, L.F. DiMauro, and C.D. Lin, Nature {\bf 483}, 194 (2012).
\bibitem{JensDiff} M.G. Pullen, et al., Nat. Comm. {\bf 6}, 7262 (2015).
\bibitem{blaga} C. I. Blaga, F. Catoire, P. Colosimo, G. G. Paulus, H. G. Muller, P. Agostini, and L. F. DiMauro, Nat. Phys. {\bf 5}, 335 (2009).
\bibitem{JensMidIR1} I. Pupeza, et al. Nat. Phot. {\bf 9}, 721 (2015).
\bibitem{dura_13} J. Dura, et al., Sci. Rep. {\bf 3}, 2675 (2013).
\bibitem{JensMidIR2} B. Wolter, et al., Phys. Rev. X {\bf 5}, 021034 (2015).
\bibitem{faisal} F.H. Faisal, J. Phys. B {\bf 6}, L89 (1973).
\bibitem{reiss} H.R. Reiss, Phys. Rev. A {\bf 22}, 1786 (1980).
\bibitem{bandrauk2} A.D. Bandrauk and S. Chelkowski, Phys. Rev. Lett. {\bf 84}, 3562 (2000).
\bibitem{landsman2013} A.S. Landsman, C. Hofmann, A.N. Pfeiffer, C. Cirelli, and U. Keller, Phys. Rev. Lett. {\bf 111}, 263001 (2013).
\bibitem{brabec} T. Brabec, M.Y. Ivanov, and P.B. Corkum, Phys. Rev. A {\bf 54}, R2551 (1996).
\bibitem{quan} W. Quan, et al. Phys. Rev. Lett. {\bf 103}, 093001 (2009).
\bibitem{wu} C.Y. Wu, Y.D. Yang, Y.Q. Liu, and Q.H. Gong, Phys. Rev. Lett. {\bf 109}, 043001 (2012).
\bibitem{wolter_14} B. Wolter, et al., Phys. Rev. A {\bf 90}, 063424 (2014).
\bibitem{moller_14} M. M{\"o}ller, F. Meyer, A. M. Sayler, G. G. Paulus, M. F. Kling, B. E. Schmidt, W. Becker, and D. B. Milo\u{s}evi\'{c}, Phys. Rev. A {\bf 90}, 023412 (2014).
\bibitem{liu} C. Liu and K.Z. Hatsagortsyan, Phys. Rev. Lett. {\bf 105}, 113003 (2010).
\bibitem{yan} T.M. Yan, S.V. Popruzhenko, M.J.J. Vrakking and D. Bauer, Phys. Rev. Lett. {\bf 105}, 253002 (2010).
\bibitem{xia} Q.Z. Xia, D.F. Ye, L.B. Fu, X.Y. Han, and J. Liu, Sci. Rep. {\bf 5}, 11473 (2015).
\bibitem{rost} A. Kaestner, U. Saalmann, and J.M. Rost, Phys. Rev. Lett. {\bf 108}, 033201 (2012).
\bibitem{peternature} M. Krueger, M. Schenk, and P. Hommelhoff, Nature {\bf 475}, 78 (2011).
\bibitem{kling} F. S\"{u}{\ss}mann and M. F. Kling, Phys. Rev. B {\bf 84}, 121406(R) (2011).
\bibitem{peterprl2006}  P. Hommelhoff, Y. Sortais, A. Aghajani-Talesh, and M.A. Kasevich, Phys. Rev. Lett. {\bf 96}, 077401 (2006).
\bibitem{peterprl2010} M. Schenk, M. Krueger, and P. Hommelhoff, Phys. Rev. Lett. {\bf 105}, 257601 (2010).
\bibitem{peterjpbreview} M. Kr\"uger, M. Schenk, M. F\"orster, and .P Hommelhoff, J. Phys. B {\bf 45}, 074006 (2012).
\bibitem {ropers} G. Herink, D.R. Solli, M. Gulde, and C. Ropers, Nature {\bf 483}, 190 (2012).
\bibitem{z11} S. Zherebtsov \textit{et al.}, Nat. Phys. {\bf 7}, 656 (2011).
\bibitem{sussmann2015field} F. S{\"u}{\ss}mann, et al., Nat. Comm. {\bf 6}, 7944 (2015).
\bibitem{forg2016attosecond} B. F{\"o}rg, et al., Nat. Comm. {\bf 7}, 11717 (2016).
\bibitem{kimNature} S. Kim, J. Jin, Y.-J. Kim, I.-Y. Park, Y. Kim, and S.-W. Kim, Nature {\bf 453}, 757 (2008).
\bibitem{sivis} M. Sivis, M. Duwe, B. Abel and C. Ropers, Nat. Phys. {\bf 9}, 304 (2013).
\bibitem{husakou} A. Husakou, S.-J. Im, and J. Herrmann, Phys. Rev. A {\bf 83}, 043839 (2011). 
\bibitem{ciappi2012} M. F. Ciappina, J. Biegert, R. Quidant, and M. Lewenstein, Phys. Rev. A {\bf 85}, 033828 (2012).
\bibitem{choi}  S. Choi, M. F. Ciappina, J. A. P{\'e}rez-Hern{\'a}ndez, A. S. Landsman, Y.-J. Kim, S. C. Kim, and D. Kim, Phys. Rev. A {\bf 93}, 021405, (2016).
\bibitem{yavuz} I. Yavuz, Phys. Rev. A {\bf 85}, 013416 (2012).
\bibitem{ciappi_opt} M. F. Ciappina, S. S. Acimovic, T. Shaaran, J. Biegert, R. Quidant, and M. Lewenstein, Opt. Exp. {\bf 20}, 26261 (2012).
\bibitem{ciappiati1d}M. F. Ciappina, J. A. P{\'e}rez-Hern{\'a}ndez, T. Shaaran, J. Biegert, R. Quidant, and M. Lewenstein, Phys. Rev. A {\bf 86}, 023413 (2012).
\bibitem{tahirsfa}T. Shaaran, M. F. Ciappina, and M. Lewenstein, Phys. Rev. A {\bf 86}, 023408 (2012).
\bibitem{tahirJMO}T. Shaaran, M. F. Ciappina, and M. Lewenstein, J. Mod. Opt. {\bf 59}, 1634 (2012).
  \bibitem{LPL} M. F. Ciappina, T. Shaaran, R. Guichard, J. A. P\'erez-Hern\'andez, L. Roso, M. Arnold, T. Siegel, A. Za\"ir, and M. Lewenstein, Las. Phys. Lett. {\bf 10}, 105302 (2013).
\bibitem{PRA_} M. F. Ciappina, J. A. P\'erez-Hern\'andez, T. Shaaran, L. Roso, and M. Lewenstein, Phys. Rev. A  {\bf 87}, 063833 (2013).
\bibitem{lewenstein1995rings} M. Lewenstein, K. C. Kulander, K. J. Schafer, and P. H. Bucksbaum, Phys. Rev. A  {\bf 51}, 1495 (1995).
\bibitem{salieres2001feynman} P. Sali{\`e}res, et al., Science {\bf 292}, 902 (2001).
\bibitem{suarez2015above} N. Su{\'a}rez, A. Chac{\'o}n, M. F. Ciappina, J. Biegert, and M. Lewenstein, Phys. Rev. A  {\bf 92}, 063421 (2015).
\bibitem{tong} X. M. Tong and C. D. Lin, J. Phys. B: At. Mol. Opt. Phys. {\bf 38}, 2593 (2005).
 \bibitem{schaferwop1} K. J. Schafer and K. C. Kulander, Phys. Rev. A  {\bf 42}, 5794 (1990).
 \bibitem{schaferwop} K. J. Schafer, Comput. Phys. Commun. {\bf 63}, 427 (1991).
\bibitem{schaferwop2} K. J. Schafer, Numerical Methods in Strong Field Physics, in Strong Field Laser Physics, ed. T. Brabec, Springer Series in Optical Sciences (Springer, Berlin, 2008). 
\bibitem{schaf93A} K. Schafer, B. Yang, L. DiMauro, and K. Kulander, Phys. Rev. Lett. {\bf 70}, 1599 (1993).
\bibitem{corku93A} P. B. Corkum,  Phys. Rev. Lett. {\bf 71}, 1994 (1993).
\bibitem{ammosov1986tunnel} M.V. Ammosov, N.B. Delone, and V.P. Krainov, Sov. Phys.-JETP {\bf 64}, 1191 (1986). 
\bibitem{delone1991energy} N. B. Delone and V. P. Krainov, J. Opt. Soc. Am. B {\bf 49}, 6 (1991).
\bibitem{arissian} L. Arissian, C. Smeenk, F. Turner, C. Trallero, A. V. Sokolov, D. M. Villeneuve, A. Staudte, and P. B. Corkum, Phys. Rev. Lett. {\bf 105}, 133002 (2010).
\bibitem{rydberg} A.S. Landsman, A.N. Pfeiffer, C. Hofmann, M. Smolarski, C. Cirelli, and U. Keller, New J. Phys. {\bf 15}, 013001 (2013).
\bibitem{JPhysBTime} A.S. Landsman and U. Keller, J. Phys. B {\bf 47}, 204024 (2014).
\bibitem{eichmann} T. Nubbermeyer, K. Gorling, A. Saenz, U. Eichmann, and W. Sandner, Phys. Rev. Lett. {\bf 101} 233001 (2008).
\bibitem{hofmann2013comparison} C. Hofmann, A. S. Landsman, C. Cirelli, A. N. Pfeiffer, and U. Keller, J.  Phys. B {\bf46}, 125601 (2013).
\bibitem{pfeiffer2012attoclock} A. N. Pfeiffer, C. Cirelli , M. Smolarski, D. Dimitrovski, M. Abu-samha, L. B. Madsen, and U. Keller, Nat. Phys. {\bf 8}, 76 (2012).
\bibitem{Cornelia2} C. Hofmann, A.S. Landsman, A. Zielinski, C. Cirelli, T. Zimmermann, A. Scrinzi, and U. Keller, Phys. Rev. A {\bf 90}, 043406 (2014). 
\bibitem{Cornelia3} C. Hofmann, T. Zimmermann, A. Zielinski, and A.S. Landsman, New J. Phys {\bf 18}, 043011 (2016).
\bibitem{yang1993intensity} B. Yang, K. J. Schafer, B. Walker, K. C. Kulander, P. Agostini, and L. F. DiMauro, Phys. Rev. Lett. {\bf 71}, 3770 (1993).
\bibitem{paulus_94} G. G. Paulus, W. Nicklich, H. Xu, P. Lambropoulos, and H. Walther, Phys. Rev. Lett. {\bf72}, 2851 (1994).
\bibitem{yudin2001nonadiabatic} G. L. Yudin and M. Y. Ivanov, Phys. Rev. A {\bf 64}, 013409 (2001).
\bibitem{paulusCEP} G.G. Paulus, F. Lindner, H. Walther, A. Baltuska, E. Goulielmakis, M. Lezius, and F. Krausz, Phys. Rev. Lett. {\bf 91},  253004 (2003).
\bibitem{milosevic_03} D. B. Milo\u{s}evi\'{c}, G. G. Paulus, and W. Becker, Opt. Exp. {\bf11}, 1418 (2003).
\bibitem{carpetPRL} Ph. A. Korneev, et al, Phys. Rev. Lett. {\bf 108}, 223601 (2012).
\bibitem{wirth} A. Wirth, et al., Science {\bf 334}, 195 (2011).
\bibitem{fill} E. Fill, L. Veisz, A. Apolonski, and F. Krausz, New J. Phys. {\bf 8}, 272 (2006).
\bibitem{baum} C. Kealhofer, W. Schneider, D. Ehberger, A. Ryabov, F. Krausz, and P. Baum, Science {\bf 352}, 6284 (2016).

\end{thebibliography}

\end{document}